\documentclass[prl, 10pt, twocolumn, floatfix, superscriptaddress]{revtex4-1}
\usepackage{amsmath} 
\usepackage{graphicx} 
\usepackage{verbatim}
\usepackage{color} 
\usepackage{subfigure} 
\usepackage{hyperref} 
\usepackage{sidecap}
\usepackage{tikz}
\usepackage{amsmath,bm}
\usepackage{mathtools}

\begin{document}

\title{Stationary states of trapped spin-orbit-coupled Bose--Einstein condensates}
\author{E. Ruokokoski}
\author{J. A. M. Huhtam\"aki}
\affiliation{COMP Centre of Excellence, Department of Applied Physics, Aalto University,
  P.O.~Box 13500, FI-00076 AALTO, Finland}
\author{M. M\"ott\"onen$^{1,}$}
\affiliation{Low Temperature Laboratory (OVLL), Aalto University, P.O.~Box 13500, FI-00076 AALTO, Finland}


\begin{abstract}
We numerically investigate low-energy stationary states of pseudospin-1 Bose--Einstein condensates in the presence of Rashba--Dresselhaus-type 
spin-orbit coupling. We show that for experimentally feasible parameters and strong spin-orbit coupling, 
the ground state is a square vortex lattice irrespective of the nature of the spin-dependent interactions. For weak spin-orbit coupling, 
the lowest-energy state may host a single vortex. Furthermore, we analytically derive constraints 
that explain why certain stationary states do not emerge as ground states. Importantly, we show that the distinct stationary states can 
be observed experimentally by standard time-of-flight spin-independent absorption imaging.
\end{abstract}

\maketitle

\emph{Introduction---}Ultracold atomic gases have proven to be excellent systems to emulate various phenomena 
arising in condensed matter 
and high-energy physics. However, due to the charge neutrality of the constituent atoms, no Lorentz force acts on these 
systems in the presence of an electromagnetic field. This apparently limits the use of these systems in simulating 
phenomena arising from the coupling between a matter field and a gauge field. Therefore, methods 
to generate effective gauge potentials for ultracold atoms have been actively 
studied in recent years~\cite{Osterloh:2005, Lin_PRL:2009, Liu:2007, Zhu:2006, Dalibard:2011, Fu:2011}. One way to create such an artificial gauge field, is to 
couple the atoms with spatially varying laser fields~\cite{Juzeliunas:2006, Lin_Nature:2009}. 
This scheme is advantageous since it can be used to create both 
Abelian and non-Abelian gauge potentials~\cite{Ruseckas:2005, Osterloh:2005}, and the resulting field may be 
controlled and shaped by adjusting the laser beams~\cite{Juzeliunas:2005}.

These techniques can be used to create an artificial gauge potential that mimics the vector potential arising from the
spin-orbit coupling (SOC), i.e., the interaction that couples the spin and momentum degrees of freedom~\cite{Stanescu:2008, Lin_Nature:2009,
Juzeliunas:2010, Ho:2011, Wu:2011, Anderson:2011, Liu:2012, Li:2012}. 
A special case of SOC is the Rashba--Dresselhaus (RD) coupling which is actively studied due to 
its relevance in spintronics~\cite{Datta:1990, Das:1990, Ganichev:2004}. 
In RD coupling, the vector potential is proportional to the spin-$1/2$ operator of a particle within a plane. 
In Ref.~\cite{Juzeliunas:2010}, it was proposed that RD-type SOC could be generalized for spins larger than $1/2$ 
in cold atomic gases using the so-called $N$-pod setup, where $N$ laser beams are used to couple $N$ internal atomic ground states 
to a common auxiliary state. A tetrapod setup ($N=4$) was suggested to generate RD-type coupling where the  vector potential is proportional
to a spin-$1$ operator within a plane. The RD coupling has been shown to generate non-trivial 
structures in spin-$1/2$, spin-$1$, and spin-$2$ Bose--Einstein condensates (BECs)~\cite{Wang:2010, HuHui:2012, Sinha:2012, Ozawa:2012, Xu:2012, 
Kawakami:2011, Xu:2011, Yip:2011}.

In the homogeneous RD-coupled BEC, the solutions of the single-particle Hamiltonian are plane waves. These solutions provide 
insight into the stationary states of a trapped system, which can be approximated by 
superpositions of different number of the degenerate single-particle solutions. In Ref.~\cite{Wang:2010}, it was stated that the ground state of 
the trapped RD coupled condensate has two possible phases in the spin-1 case, namely the plane-wave (PW) and the standing-wave (SW) phases. 
In the PW phase the ground state of the condensate can be approximated by a single plane wave, whereas in the SW phase each spinor 
component consists of two counter-propagating coherent plane waves forming a standing wave. In Ref.~\cite{Xu:2012}, symmetry 
properties of the system were used to classify the ground-states of a trapped BEC with strong RD coupling. It was predicted that
in addition to the PW and SW states, also exotic lattice states, namely, 
the triangular-lattice state and the square-lattice (SL) state emerge as the ground states of the spin-$1$ condensate.  
These states are superpositions of three and four plane waves, respectively, and they are invariant under simultaneous discrete spin and
space rotations about the $z$-axis. Such vortex lattices cannot be created by rotating the condensate as the total angular momentum of 
these states vanishes. The SL state has also been predicted to occur in rapidly quenched spin-orbit coupled Bose gases~\cite{Su:2011}. 
Furthermore, in Ref.~\cite{Xu:2012}, states preserving the combined SO(2) spin-space rotational symmetry of the Hamiltonian were 
found to be ground states in some regions of the parameter space. These states can be approximated by an infinite number of
degenerate single-particle solutions. 

In this Rapid Communication, we analyze the energetics of stationary states arising from the RD-type SOC in an optically trapped pseudospin-1 BEC. 
In Refs.~\cite{Wang:2010} and~\cite{Xu:2012}, exotic ground states were predicted to emerge but only a few discrete values of parameters 
were considered. In our analysis, we concentrate on the effect of the SOC strength on the energies of the stationary states for realistic 
density-density coupling strengths and determine the ground state of the condensate in different regions of the parameter space. 
The SL state is found to emerge as the ground state for strong SOC, irrespective of the spin-dependent interactions.
Starting from the homogeneous approximation, we analytically derive constraints for the possible ground states and neglect the states that are found 
to be energetically unfavorable. These unfavorable states include the triangular lattice, described above, for the reasons 
we will explicate below. Furthermore, we propose a robust method to observe the exotic stationary states by imaging only the particle density in 
time-of-flight experiments.

\emph{Theory---}In the tetrapod setup described in Ref.~\cite{Juzeliunas:2010}, 
four laser beams are used to couple four internal atomic ground states to a common auxiliary state. This coupling 
gives rise to three degenerate dark states, that is, zero-energy eigenstates of the atom-light Hamiltonian that 
are superpositions of the four atomic ground states. The condensed atoms reside in the dark states which
play the role of internal pseudospin degrees of freedom~\cite{Stanescu:2008}, and the order parameter of the system takes the 
form of a three-component spinor denoted by $\Psi=\left(\Psi_1,\,\Psi_0,\,\Psi_{-1}\right)$. 

The stationary states of the system are solved from the time-independent Gross--Pitaevskii (GP) equation
\begin{equation}\label{eq:GP}
\mathcal{H}[\Psi]\Psi(\boldsymbol{r})=
\mu\Psi(\boldsymbol{r}),
\end{equation}
where the effective Hamiltonian of the system is given by
\begin{equation}\label{eq:Hamilt1}
\mathcal{H}\left[\Psi\right]=\frac{1}{2m}\left(\frac{\hbar}{i}\nabla-\alpha\boldsymbol{A}\right)^{2}+\Theta+
V(\boldsymbol{r})+c_{0}\Psi^{\dagger}(\boldsymbol{r})\Psi(\boldsymbol{r}).
\end{equation}
Here $m$ is the atomic mass,
$V$ is the optical trapping potential, $\alpha$ is the SOC strength, 
and $c_0$ is the density-density coupling constant.
The vector $\boldsymbol{A}$ represents the non-Abelian gauge potential and the matrix $\Theta$ is an effective scalar potential. In the tetrapod setup 
$\Theta=\alpha^2\mathcal{F}^2_z/(2m)$ and the vector potential is of the Rashba--Dresselhaus form 
$\bm{A}=\mathcal{F}_x\hat{\bm{e}}_x+\mathcal{F}_y\hat{\bm{e}}_y$ ~\cite{Juzeliunas:2010}.
With this vector potential and assuming that the optical trapping potential $V$ is cylindrically symmetric, the Hamiltonian in Eq.~(\ref{eq:Hamilt1}) 
is invariant under the combined spin and spatial rotation about the
$z$-axis $\hat{\bm{R}}=e^{i\gamma\hat{\bm{F}}_z + i\gamma\hat{\bm{L}}_z}$. 

The simple Hamiltonian in Eq.~(\ref{eq:Hamilt1}) yields a good description of the system in the case that the four internal atomic states 
coupled to the auxiliary state are chosen such that the density-density coupling is roughly independent of the internal state. For example, if the states are 
chosen from the $F=2$ manifold of $^{87}\textrm{Rb}$, the atom-atom interaction Hamiltonian 
is a sum of the density-density, spin-spin, and spin-singlet pairing interaction terms~\cite{Ciobanu:2000}. The magnitudes of the latter two terms are 
of order $1\%$ of the first one~\cite{Widera:2006}. Hence, we obtain the Hamiltonian in Eq.~(\ref{eq:Hamilt1}) by keeping only the density-density term 
which retains its form also in the dark state basis. Counting on the possibility that an analogous spin-orbit interaction is realized in a genuine spin-1 system, we also perform calculations 
in the presence of a spin-spin coupling term $\mathcal{H}_{\textrm{ss}}=c_2\Psi^{\dagger}(\boldsymbol{r})\boldsymbol{\mathcal{F}}\Psi(\boldsymbol{r})\cdot\boldsymbol
{\mathcal{F}}$~\cite{Ohmi:1998, Ho}, where $c_2$ is the spin-spin coupling constant and $\boldsymbol{\mathcal{F}}=\left(\mathcal{F}_{x},\,\mathcal{F}_{y},\,
\mathcal{F}_{z}\right)^{T}$ is a vector of spin-1 matrices. We consider both the ferromagnetic ($c_2<0$) and the antiferromagnetic ($c_2>0$) interactions.

\begin{figure*}
\includegraphics[width=1\textwidth]{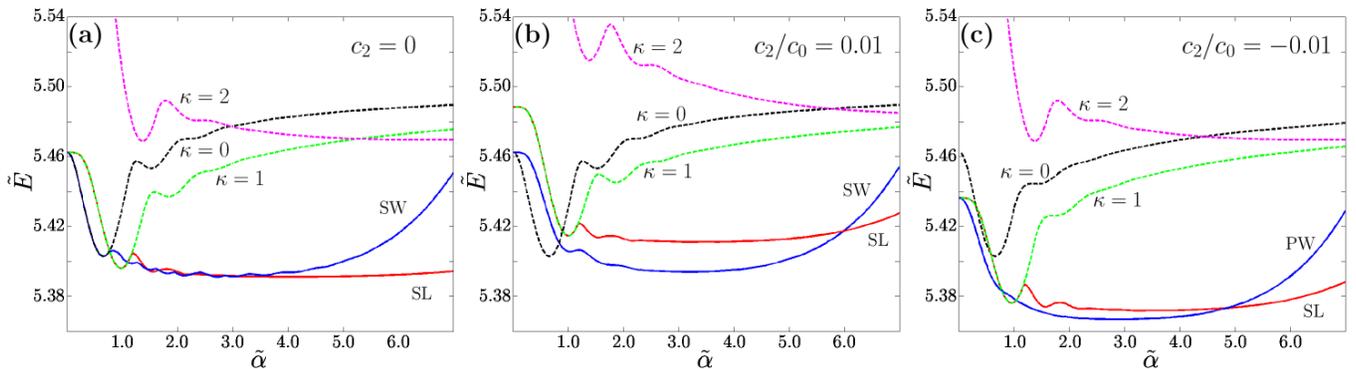}
\caption{\label{fig:g_curves}(Color online) Energies of low-energy stationary states for effectively two-dimensional 
condensates as functions of the SOC strength. Panels (a)--(c) represent the pseudospin case  $c_2=0$, 
the antiferromagnetic case ($c_2/c_0=0.01$), and the ferromagnetic case ($c_2/c_0=-0.01$), respectively. 
Here we take $\tilde{c}_0=200$. The various stationary states can be approximated by superpositions of degenerate plane-wave solutions of the single-particle Hamiltonian.
The plane-wave (PW) state can be approximated by a single plane wave, whereas the standing-wave (SW) state
corresponds to a superposition of two coherent counter-propagating plane waves. Furthermore, the square-lattice (SL) state is formed by two pairs of counter-propagating plane waves. 
The dashed lines correspond to cylindrically symmetric states with different values of $\kappa$ in Eq.~(\ref{eq:ansatz}). Here, $0.5\,\tilde\alpha^2$ has been subtracted from 
the energies $\tilde{E}$ for clarity.}
\end{figure*}

Significant understanding of our results can be gained by considering 
the homogeneous RD-coupled BEC. The minimum-energy eigenstate of the single-particle Hamiltonian,
$\mathcal{H}_{1}=\left(-\hbar^2\nabla^2+2i\hbar\alpha\boldsymbol{A}\cdot\nabla+2\hbar^2\alpha^2\right)/2m$, 
is a plane wave $\Psi_{\bm{k}}=\Phi_{\bm{k}} e^{i\bm{k}\cdot\bm{r}}$~\cite{Juzeliunas:2010}, where $\bm{k}$ lies in the 
$xy$-plane, $\Phi_{\bm{k}}=\left(1,\,\sqrt{2}e^{i\theta_{\bm{k}}},\,e^{2i\theta_{\bm{k}}}\right)^{T}/2$,
and $\theta_{\bm{k}}=\arctan(k_y/k_x)$. The corresponding energy is given by
$\mathcal{E}_{k}=\hbar^2\left(k^2-2\alpha k +2\alpha^2\right)/2m$. 
Let us consider states which are superpositions of the degenerate single-particle ground states
\begin{equation}\label{eq:super}
\Psi^{(n)}=Bf\left(\bm{r}\right)\sum_{j=1}^{n}a_j\Phi_{\bm{k}_j}e^{i\bm{k}_j\cdot\bm{r}},
\end{equation}
where $B$ is a normalization constant and $f(\bm{r})$ models the presence of the trapping potential and the effect of repulsive interactions.
Thus, $\left|f\left(\bm{r}\right)\right|^2$ yields a density profile characteristic of trapped condensates.
It is straightforward to show analytically that, irrespective of $f(\bm{r})$, 
the SOC energies of the states with arbitrary $n$ are degenerate, $\mathcal{E}_{\mathrm{SOC}}=\int\Psi^{(n)\dagger}\left(2i\hbar\alpha\bm{A}\cdot\nabla 
+ 2\hbar^2\alpha^2\right)\Psi^{(n)}\textrm{d}\bm{r}/2m=\hbar^2\left(-\alpha k +\alpha^2\right)/m$. However, some of these superposition 
states give rise to density profiles which are energetically unfavorable. Hence, we require that the condensate density profile should be smooth, i.e.,
\begin{equation}\label{eq:smooth}
\nabla\left|\Psi^{(n)}\left(\bm{r}\right)/f\left(\bm{r}\right)\right|^2=0.
\end{equation}
This requirement assures that the interference of the plane waves does not cause spatial variations in the density.
We note that if the density-density coupling is very weak, the energy contribution coming from the density modulations might not be large enough 
to render the state energetically unfavorable. However, for typical experimental values of $c_0$, Eq.~(\ref{eq:smooth})
should be fulfilled. Equation~(\ref{eq:smooth}) is trivially satisfied for $n=1$, but for other odd values of $n$, the condition 
never holds, implying that these states are energetically unfavorable due to density modulations. For $n=2$ the requirement is satisfied if 
$\left|\theta_{1}-\theta_{2}\right|=\pi$, where $\theta_{j}=\arctan(k^j_{y}/k^j_{x})$. For other even superpositions, the requirement 
is satisfied only if the constants $a_j$ in Eq.~(\ref{eq:super}) satisfy the condition $a^{\ast}_{i}a_j+a^{\ast}_{j^{\prime}}a_{i^{\prime}}=0$ 
with $i,j=1,2,\ldots,n$, $\theta_{i^{\prime}(j^{\prime})}=\theta_{i(j)}+\pi$, and $i\neq j^{\prime}$, $j\neq i^{\prime}$. 
We conclude that the energetically favorable solutions consist of either a single $\bm{k}$ or one or multiple pairs $\left\{\bm{k},\,-\bm{k}\right\}$. 
In particular, the above condition implies that the counter-propagating plane waves must have equal amplitudes. When $n\rightarrow\infty$, the ansatz in Eq.~(\ref{eq:super})
gives a cylindrically symmetric state and the order parameter $\Psi$ in the cylindrical coordinates $(\rho,\phi, \,z)$ reads
\begin{equation}\label{eq:ansatz}
\Psi\left(\rho,\,\phi,\,z\right)=\left(\!\begin{array}{c}
\Psi_1\left(\rho,\,z\right)e^{i\left(\kappa-1\right)\phi}\\
\Psi_0\left(\rho,\,z\right)e^{i\kappa\phi}\\
\Psi_{-1}\left(\rho,\,z\right)e^{i\left(\kappa+1\right)\phi}
\end{array}\!\right),
\end{equation}
describing a single spin vortex. Time reversal symmetry implies that for each $\kappa\ne 0$ there is a degenerate state with $-\kappa$, and thus we consider only $\kappa\ge 0$.

\emph{Numerical results---}
The stationary states are found by solving the time-independent GP equation (\ref{eq:GP}). In order to obtain the excited states, we solve the time-independent 
GP equation enforcing symmetries characteristic of each state. For the cylindrically symmetric stationary states the form of the order parameter is fixed by Eq.~(\ref{eq:ansatz}).
We carried out most of the computations for a pancake-shaped 
condensate and assumed Gaussian profile in the axial direction. A truly three-dimensional system was simulated for various values of $\alpha$ and the coupling constant $c_2$  and the obtained ground states agree with 
the ones obtained in the effectively two-dimensional case. We measure length in units of the radial harmonic oscillator length $a_r=\sqrt{\hbar/(m\omega_r)}$
and energy in units $\hbar\omega_r$. The dimensionless coupling constants for the effectively two-dimensional case are given by 
$\tilde{c}_0\approx\sqrt{8\pi}N a/a_z$, and $\tilde{\alpha}=a_r\alpha$. 
Here, $a$ is the vacuum $s$-wave scattering length and $a_z=\sqrt{\hbar/(m\omega_z)}$ is the axial harmonic oscillator length.
The exact form of $\tilde{c}_0$ is setup dependent.
\begin{figure}
\includegraphics[width=1\linewidth]{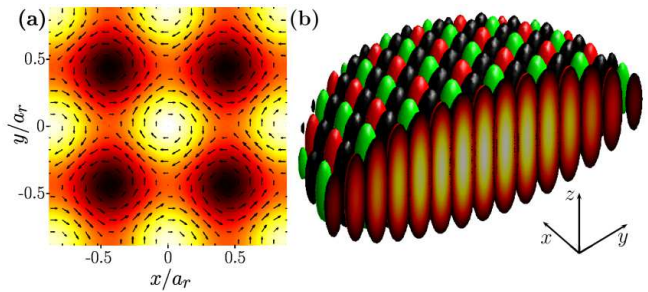}
\caption{\label{fig:mag3D}(Color online) (a) The magnetization of the square-lattice state. The arrows represent the projection of the 
magnetization to the $xy$-plane and the $z$-component of the magnetization is illustrated with the colormap. Black color corresponds to small values and white color to large values. 
(b) The density isosurfaces of the spinor components in the square-lattice state. The 
green color corresponds to $\left|\Psi_{-1}\right|^2$, the black color to $\left|\Psi_{0}\right|^2$ and the 
red one to $\left|\Psi_{1}\right|^2$. The field of view is $6.6 \times 3.3 \times 2.4\,a_r^3$ and the isosurfaces correspond 
to the density $0.02\, N/a_r^3$. The density isosurfaces are capped with density colormap where 
the lightest color corresponds to the largest density. The dimensionless value of the SOC strength was taken to be $\tilde{\alpha}=5.0$.}
\end{figure}

\begin{figure}
\includegraphics[width=1\linewidth]{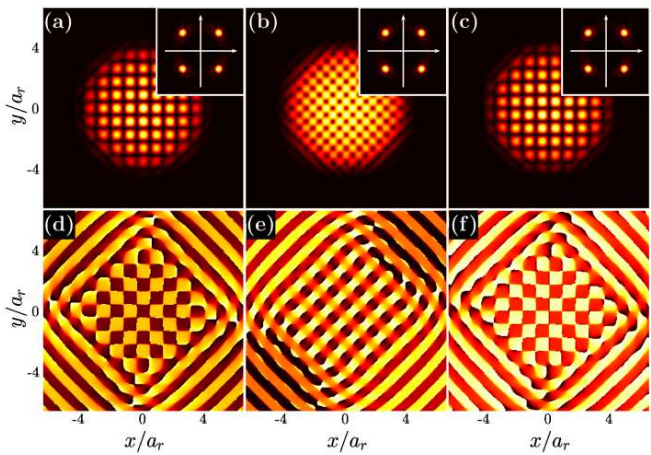} 
\caption{\label{fig:hila_dens}(Color online) Spinor components 
in the square-lattice state. The orientation of the axes in all the panels is the same. Panels (a)--(c) represent the amplitudes and panels 
(d)--(f) the phases of $\Psi_1$, $\Psi_0$, $\Psi_{-1}$, respectively. The insets in panels (a)--(c) represent the absolute value of the Fourier transform for the 
corresponding spinor component. The horizontal and the vertical axes on the insets correspond to $k_x$ and $k_y$, respectively. The values of the 
coupling constants were taken to be $\tilde{c}_0=200$, $c_2=0$, and $\tilde{\alpha}=5.0$.}
\end{figure}

Figure~\ref{fig:g_curves} shows the energies of various stationary states 
as functions of the SOC strength $\alpha$. 
For the pseudospin case with $c_2=0$ the SW and SL states are nearly degenerate for the intermediate values of $\alpha$, whereas for 
the antiferromagnetic case with $c_2>0$, the SW state is found to be the ground 
state of the condensate. For the ferromagnetic case with $c_2<0$, the PW state emerges 
for small and intermediate values of $\alpha$. We observe that in all three cases, there is a region where the ground state of 
the condensate is a cylindrically symmetric vortex state, corresponding to $n\to\infty$ in Eq.~(\ref{eq:super}). 
For the ferromagnetic case, the ground state hosts a Mermin--Ho vortex with $\kappa=1$ in Eq.~(\ref{eq:ansatz}), 
whereas for the antiferromagnetic case the $\kappa=0$ state emerges as the ground state for small values of $\alpha$. 
In the absence of spin-spin coupling, both $\kappa=0$ and $\kappa=1$ vortices appear as ground states.
Furthermore, for large SOC strengths the lowest-energy state is, irrespective of the value of $c_2$, 
a square lattice (SL) state, which can be approximated by putting $n=4$, 
$\left|a_1\right|^2=\left|a_2\right|^2=\left|a_3\right|^2=\left|a_4\right|^2$, and $\theta_{j}=j\pi/2$ in Eq.~(\ref{eq:super}). 
We found two of these lattice states: one in which the $\Psi_0$ component has a density minimum in the middle of the trap and a phase 
singularity on the $z$-axis, and another, shifted so that $\Psi_0$ has a density maximum at the center and no phase singularity on the 
$z$-axis. These states are nearly degenerate and their energetics are qualitatively the same. Hence, we only present results for the former one. 
The magnetization and the three-dimensional particle density for the SL state are shown in Fig.~\ref{fig:mag3D}. We observe two kinds of spin 
vortices: polar core vortices and vortices with a ferromagnetic core polarized alternately in $z$ and $-z$ directions, corresponding to 
Mermin--Ho vortices and antivortices. This kind of alternating vortex structure cannot be created by rotating the condensate as the total angular momentum 
of the state vanishes. The amplitude, complex phase, and the Fourier transform of each spinor component in the SL state are 
shown in Fig.~\ref{fig:hila_dens}.

The oscillations in the SL state energy curves in Figs.~\ref{fig:g_curves}(a)--(c) are caused by the increase in the vortex density
as the SOC strength increases. On the other hand, the oscillations in the curves for the SW state and the radially symmetric states are caused by 
the increasing number of nodes in the density profiles of the spinor components. We also investigated the effect of the density-density coupling on 
the energetics of the SL state. We observed that increasing $c_0$ increases the critical value of $\alpha$ for which the SL state becomes the 
ground state of the system. Hence, for strongly interacting BECs, strong SOC is required to render the SL state the ground state.
 
\begin{figure}
\begin{minipage}[!b]{1\linewidth}
\includegraphics[width=1\textwidth]{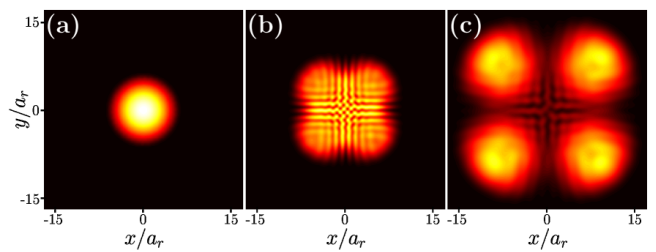}
\caption{\label{fig:aika}(Color online) Three-dimensional dynamics of the square-lattice state after the removal of the SOC and the optical 
trapping in the $x$- and $y$-directions. The density is integrated in the $z$-direction. Panels (a)--(c) represent 
the density profile at $t=0,\,t=1/\omega_r\textrm{ and }t=2/\omega_r$, respectively. The maxima of the colormap in panels (a)--(c) 
are $0.03$, $0.0080$ and $0.0035$ in units of $N/a_r^3$. The
values of the coupling constants were taken to be $\tilde{c}_0=1000$, $c_2=0$, $\tilde{\alpha}=5.0$.
}
\end{minipage}
\end{figure}

Finally, we demonstrate that the exotic stationary states of the spin-orbit-coupled condensates can be observed by 
time-of-flight experiments. Since the stationary states can be approximated by a single plane wave or superpositions of 
standing waves, they are characterized by either a single $\bm{k}$ or pairs $\left\{\bm{k},\,-\bm{k}\right\}$. 
Hence, when such a state is left to evolve in time after instantaneously removing the SOC and the optical trapping in the $x$-and $y$-directions, 
one would expect that the condensate density begins to move in the directions 
specified by the $\bm{k}$-vectors. Figure~\ref{fig:aika} shows the three-dimensional dynamics for the SL state after the 
$xy$-trap and the SOC were switched off. We observe that the condensate 
density separates into four equal portions moving in the four directions dictated by the 
$\bm{k}$-vectors. We also computed the temporal evolution for the SW state and observed that it separates into two segments moving 
in opposite directions.  For the PW state, the condensate density moves in the direction specified by the single $\bm{k}$, whereas the 
cylindrically symmetric states of Eq.~(\ref{eq:ansatz}) expand radially. Thus the states can be observed from the total particle density by a standard absorption imaging technique, regardless of 
whether the states are obtained using a genuine spin-1 condensate or a pseudospin-1 condensate. 
Furthermore, measurement of the separation of the interference fringes visible in Fig.~\ref{fig:aika}(b) yields 
information on $\left|\bm{k}\right|.$

\emph{Conclusions---}
We have computed the energies of various stationary states of trapped Rashba--Dresselhaus coupled pseudospin-1 BECs as functions of the spin-orbit coupling strength $\alpha$.
Our results indicate that for weak spin-orbit coupling, states with a single vortex can emerge as ground states. With intermediate values of $\alpha$,
the plane-wave and standing-wave states were found to be the ground states for the cases $c_2<0$ and $c_2>0$, respectively. For strong spin-orbit coupling, 
the exotic square lattice is the ground state irrespective of the spin-spin coupling strength. This indicates that the emergence of the SL state 
as the ground state is not affected by the setup-dependent spin-spin coupling term. Importantly, we suggested 
a robust method to observe these states by imaging the total particle density of the condensate in typical time-of-flight experiments.
\begin{acknowledgments}
We thank Pekko Kuopanportti, P\"aivi T\"orm\"a, and Ville Pietil\"a for useful discussions and constructive comments.
This research has been supported by the Finnish Doctoral Programme in Computational Sciences (FICS), Emil Aaltonen Foundation, and the Academy of Finland 
through its Centres of Excellence Program (grant No. 251748) and GAIDIA project (grant No. 141015). CSC -- IT Center for Science Ltd. is acknowledged for computing resources.
\end{acknowledgments}
\bibliography{so}
\bibliographystyle{apsrev4-1}

\end{document}